\documentclass[a4paper,12pt]{article}
\usepackage{infnprep}
\usepackage{multirow}
\usepackage{mathtools}
\usepackage{bigstrut}
\usepackage[colorlinks]{hyperref}
\hypersetup{ colorlinks=true,  linkcolor=blue,  filecolor=magenta,       urlcolor=blue, citecolor=black}
\RequirePackage{mathptmx}      % use Times fonts if available on your TeX system
\RequirePackage{flushend}
\usepackage[bottom]{footmisc}
\usepackage{graphicx}
\usepackage[italian]{babel}

\usepackage[top=2.5cm, bottom=3cm, left=3cm, right=3cm]{geometry}%added now nov12
\newcommand{\Header}{
  \resizebox{15cm}{!}{
  \begin{tabular}{rl}
  \includegraphics[width=5cm, trim={50 100 0 0}]{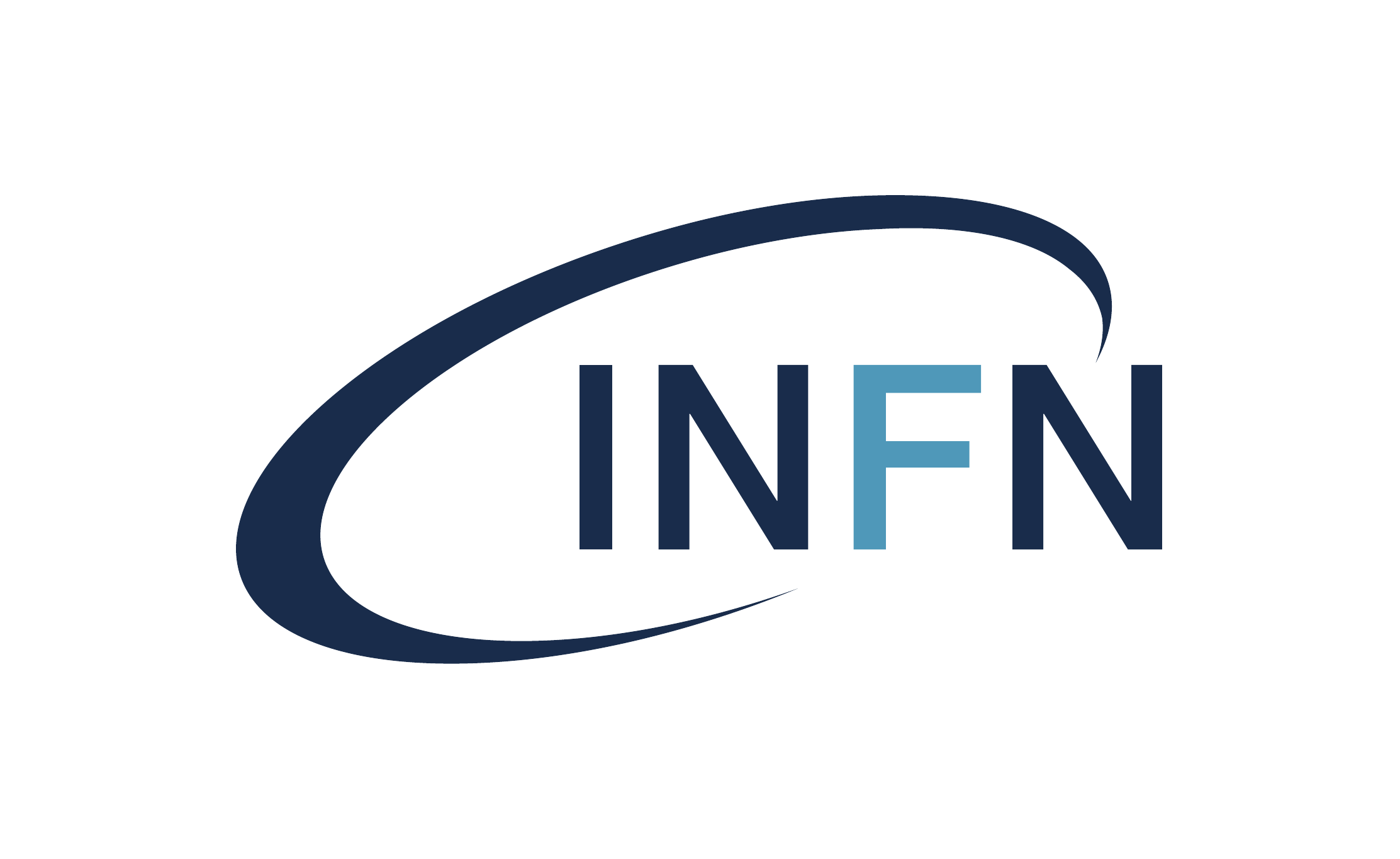} & {\LARGE\sffamily ISTITUTO NAZIONALE DI FISICA NUCLEARE}\\
      \\
  \end{tabular}
  }
\begin{center}
      {\large\sffamily Laboratori Nazionali di Frascati}\\
\end{center}
    \renewcommand{\arraystretch}{1}
\vskip 0.5cm
\rule{15.0cm}{0.09mm}
\vskip 1.5cm
  \begin{flushright}
      MIT-CTP/5254 \\
      {\underline{\bf INFN-20-17/LNF}}\\    % insert here the preprint number
      {\small\bf  11 November 2020} \\  
  \end{flushright}
  }
\begin{document}
\begin{titlepage}
\title
  {\Header {\Large \bf About soft photon  resummation}
  }
  \author{Giulia Pancheri$^1$ and Yogendra N. Srivastava$^2$\\
  {\it ${}^{1)}$INFN, Laboratori Nazionali di Frascati, P.O. Box 13,
I-00044 Frascati, Italy}\\
{\it ${}^{2)}$Dipartimento di Fisica e Geologia, Universita' di Perugia, Perugia, Italy}}
   
   \maketitle
%%%%\begin{document}
%\vskip 1 cm
%\begin{center}
%{\LARGE About soft photon  resummation} 
%\begin{flushright}
%\hspace{12cm}
%MIT-CTP/5254
%\end{flushright}
%\vskip 1cm
%Resummation and Bruno Touschek's contribution in Quantum ElectroDynamics}\\
%\author {
%%%%G. Pancheri\\INFN Frascati National Laboratories, Frascati, Italy, I00044\\\footnotetext{Also at Center for Theoretical Physics, MIT, Cambridge, MA , USA} and\\Y. N. SrivastavaDipartimento di Fisica e Geologia, Universita' di Perugia, Perugia, Italy
%}
%\end{center}

\par\noindent
\begin{center}
 A contribution to Earle Lomon's 90th birthday celebration.
 \end{center}
 %Large parts of this paper have been published in {\it Luisa Bonolis, Maestri e allievi nella fisica italiana del novecento, 2008 da completare}, and are reproduced here courtesy of dr. Bonolis.
 
 \vskip 1 cm
 \begin{abstract}
% \section*{A dedication} 
\baselineskip=14pt
The first time one of us  (G.P.)   encountered    Earle  was in Summer 1966,  when she was directed   to study Earle's  papers on radiative corrections to quasi-elastic electron scattering \cite{Lomon:1956,Lomon:1959}. The suggestion had come from   Bruno Touschek \cite{Amaldi:1981}, at the time   head of the theoretical physics group at the Frascati National Laboratories near Rome. About the same time, Earle  came from MIT  to   visit University of Rome and Frascati. G.P. was a young post-graduate,  who  had studied Earle's  papers and was awed by his  already  impressive scientific figure. After almost  40 years had passed, Earle visited Italy with his wife Ruth, making Frascati their  base for an extended visit of almost a month. They were housed in  what was then the laboratory hostel for foreign  visitors,  a small  villa higher up above the hill, toward the town of Frascati. Since then, we became close friends, a friendship which included both his family and ours, and which has been very important for us. In memory of that first visit and in gratitude for the many years of friendship, we will tell here a story of infrared radiative corrections to charged particle scattering, to  which Earle's papers gave an important contribution. 
\end{abstract}

\begin{flushright}
  \vskip 0.5cm
\small\it Published by \\
Laboratori Nazionali di Frascati
\end{flushright}
\end{titlepage}
\pagestyle{plain}
\setcounter{page}2
\baselineskip=17pt

%The story includes  what Bruno Touschek used to call the {\it Bond } factor and what many in the field consider Touschek's  legacy from QED to QCD. 

\section{The infrared catastrophe}
In late 1940's, one of the problems waiting to be solved in the newly emerging discipline of Quantum Electro Dynamics was that of the infrared catastrophe, namely the apparent divergence of  quasi-elastic electron scattering appearing  when calculated  order  by  order in perturbation theory. It had been known  that this difficulty was arising from the neglect of more than one soft photon emission \cite{Bloch:1937pw}. In 1949, Schwinger examined quasi-elastic electron scattering  and  showed that  the divergence arising from  the emission of a real photon in the limit of its energy going to zero,    is cancelled in the cross-section by a similarly divergent term arising from virtual photon absorption and emission \cite{Schwinger:1949}.  Schwinger calculated that emission up to a maximum resolution energy $\Delta E$ reduces the measured cross-section  by a factor $\delta (\Delta E)$,  
namely
\begin{equation}
\label{eq:schw1}
\sigma_{quasi-el }=\sigma_{el-theor} [1-\delta (\Delta E)]
\end{equation}
This corresponds to the well known feature that the emission of radiation reduces the cross-section of any scattering process among charged particles. 
%a well known 
%The occurence of radiation emission from accelerated particles   had also been   discussed before the waras a limiting factor to the energy attainable with a cyclotron \cite{Bethe:1937}. 

The problem however
was  not resolved by Eq.~(\ref{eq:schw1}), since $\lim_{\Delta E \rightarrow 0}\delta(\Delta E)=\infty$  and the cross-section would become negative. To avoid  this highly unpleasant 
%accident 
occurrence to show up in the calculated cross-section, 
%one needs 
it was found necessary to add the contribution of more and more soft photons. Schwinger put forward the ansatz  that  the single  soft photon contribution should be exponentiated, i.e.
\begin{equation}
\sigma_{quasi-el } \rightarrow \sigma_{el-theor} e^{-\delta(\Delta E)}
\label{eq:schw} 
\end{equation}
The cancellation between  real photons, for which   $k_{\mu}k^{\mu}=0$ and  virtual photons for which  $k_{\mu}k^{\mu}\ne 0$ points to the physical fact that 
the two types of photons are indistinguishable in the zero energy-momentum limit, 
%which is 
when the apparent divergence in photon emission arises. Brown and Feynman \cite{Brown:1952} noticed that real and virtual emissions are  physically related through the uncertainty principle: when a measurement is taken in  a given small time interval, the uncertainty introduced in the photon energy 
%may 
would allow the virtual photon to be detected as a real one. 

In \cite{Lomon:1956,Lomon:1959}, based on \cite{Jauch:1954}, the measured cross-section had been written as
\begin{equation}
\sigma(\Delta E,E,\theta)=b(\Delta E, \epsilon, C)\sigma_n(\epsilon,E,\theta) + \mathcal{O}( \epsilon/(E-m)) + \mathcal{O}[\alpha \ln (\epsilon/E)]^{n+1}] 
\end{equation}
with the infrared correction factor $b(\Delta E)$, showing the explicit cancellation in the exponentiated  photon spectrum,
%had been shown by   from Jauch and Rohrlich  
 given as \cite{Jauch:1954,Jauch:1955} 
%Their result  obtained -as quoted in Eq.(3) of \cite{Lomon:1956} 
\begin{eqnarray}
b(\Delta E,\epsilon,C)  = \frac{1}{2\pi} \int_o^{\Delta E}d\omega^{'}\int_{-\infty}^{+\infty} dt \ e^{-i\omega{'}t - h(\epsilon;t)}\\
=\frac{1}{2\pi} \int_o^{\Delta E} d\omega^{'}\int_{-\infty}^{+\infty} dt \ e^{-i\omega{'}t -  \alpha C
\int_o^\epsilon \frac{d\omega}{\omega}[1-e^{i\omega  t}]}
\label{eq:RJ}
\end{eqnarray}
where the  factor $C\equiv C(E,E')$ is a function of the incoming and outgoing particle momenta. 
%%%\begin{eqnarray}
%%%b(\Delta E)  = \frac{1}{2\pi} \int_o^{\Delta E}(d\omega^{'})\int_{-\infty}^{+\infty}(dt) e^{-i\omega{'}t - h(\epsilon;t)};
%%%\nonumber\\  
%%%h(\epsilon;t) = \alpha C \int_o^\epsilon \frac{d\omega}{\omega}[1-e^{i\omega  t}]};\nonumber\\
%%%C = (\frac{2}{\pi}) [\frac{tanh^{-1}\beta^{'}}{\beta^{'}} -1];\  \beta^{'} = \frac{|{\bf p}^{'}|}{E^{'}};\nonumber\\
%%%  [\epsilon\ {\rm a\ low\ energy\ cut-off}]\nonumber
%%%\label{eq:RJ}
%%%\end{eqnarray}
In the exponential at the  r.h.s. of Eq.~(\ref{eq:RJ}), the cancellation between real and virtual soft photons is evident, with the first term in the square bracket coming from  virtual photons, and  the $e^{i \omega t }$ coming  from the summation of real soft photons, each one of energy  $\omega \le \epsilon$. 
 The   problem of the upper integration limit in the exponential  was discussed by   Lomon   \cite{Lomon:1956}. He was later  inspired by Yennie and Suura's work  \cite{Yennie:1957} to write  in closed form the  expression for the  b-factor which modifies  the theoretical cross-section  as  \cite{Lomon:1959}
 \begin{eqnarray}
 \label{Lo}
 b(\Delta E;\epsilon;C) = (\frac{2}{\pi})\Big{(} cos\frac{\alpha C\pi}{2} \Big{)} 
 \Big{(} \frac{\Delta E}{\epsilon\gamma}\Big{)}^{\alpha C} \times I;\nonumber\\
 I = \int_o^\infty (\frac{d\sigma}{\sigma^{1+\alpha C}})\ sin \sigma\ exp\Big{(} (\alpha C)\ Ci[\frac{\epsilon}{\Delta E}\sigma] \Big{)};\nonumber\\ 
\gamma\ {\rm is\ the\ Euler\ constant}.\nonumber
 \end{eqnarray}
This equation shows the well known power law for the energy dependence of the correction factor. 
%We shall show in the next section, how  this closed form expression can also be obtained  through a semi-classical resummation calculation.   
The complete formulation of the problem - in all four dimensions -  was given in 1961 by  Yennie, Frautschi and Suura (YFS) \cite{Yennie:1961}. In the following,  we shall   show how  to obtain Lomon's expression, with the explicit power law and normalization factor, as well as the subsequent 4-dimensional formulation by YFS,  through the semi-classical method developed by Touschek to calculate
 %to and apply it
infrared radiative corrections to electron positron experiments \cite{Etim:1966zz,Etim:1967}.  We shall also discuss  Touschek's method in the context of a special kind of Abelian gauge theories, and under 
what conditions it could 
 %to deal with QED infrared radiative corrections 
 be extended to the presently still important problem of soft gluon re-summation in QCD.

 \section{Touschek resummation procedure}
In November 1960, Bruno Touschek prepared a {\it memo} for his colleagues at the University of Rome and Frascati Laboratories, in which he proposed  the construction of an electron-positron storage ring of c.m. energy $\sqrt{s}=3.0$ GeV, the highest 
%so far 
energy then under any sort of planning. It was a neat number,  an energy  chosen so as to include production of a pair of all the particles known to exist at the time, starting from two $\gamma$'s up to a $p {\bar p}$ or $n {\bar n}$. Ten months before,  the  scientific staff of the Frascati Laboratories had approved the construction of a smaller lower energy  collider, AdA, Anello di Accumulazione in Italian, storage ring in English. By November of the same year 1960, AdA was on the way to start functioning, the first of such type of machines in the world. Encouraged by this success,  Touschek  went on to propose building an electron positron collider  of much higher energy and  luminosity, one able to  produce interesting physics, which he called  ADONE, a better  AdA.

Both AdA and ADONE were electron (against positron) machines, but with a major difference: at ADONE's energy, electromagnetic radiation emission called for  important radiative corrections. In 1963,  he prepared  an internal Laboratory note \cite{Etim:1966zz}, 
reviewing the status of the contemporary   scientific literature on the subject  and laying the basis for the  subsequent more complete work  of \cite{Etim:1967}. In  tribute to Lomon's work, the note says: `The theoretical background for the idea here discussed can be found in the works by  Jauch, Rohlich and Lomon', and then again `Applying the ideas of Lomon to the problem of administering the radiative corrections 
to the work with ADONE \dots".

To calculate the correction factor to the measured $e^+e^-\rightarrow A {\bar A}$ cross-section, Touschek used the Bloch and Nordsieck result about  soft photon emission from a classical source \cite{Bloch:1937pw}. Bloch and Nordsieck showed that the distribution in the number of photons was given by a Poisson distribution, namely
 \begin{equation}
\label{eq:Poisson}
 P( \{n_{ \bf k }\}) =\Pi_{\bf k}
 \frac{
{\bar n}_{\bf k}^{n_{\bf k}}
 }
 {n_{\bf k}!} \rm{exp}[-{\bar n}_{\bf k}]
 \end{equation}
with $n_{\bf k}$ and  ${\bar n}_{\bf k}$   the number of  photons emitted with  momentum ${\bf k} $ and their average value. We notice that Eq.~(\ref{eq:Poisson})
%assume 
describes a discrete momentum spectrum  of the emitted photons, corresponding to quantization of the electromagnetic field in a finite box. In the following, we shall first assume that a smooth continuum limit exists. Later, in Sect.~\ref{sec:Palumbo}, we shall discuss possible subtleties with the continuum limit.      

Letting $K^\mu$ be the overall four-momentum loss due to  photon emission, one can sum on all  values of all the number of emitted photons $n_{\bf k}$, and write the probability $P(K)$ as
\begin{equation}
\label{eq:prob-4}
d^4 P(K)=\sum  P( \{n_{ \bf k }\})   \delta_{4}({\sum_{\bf k} k n_{\bf k} -K}) d^4 K
\end{equation}
where the four-dimensional $\delta$-function imposes overall energy momentum conservation and  allows to exchange the sum with the product in Eq.~(\ref{eq:prob-4}).   
One can then  write
\begin{equation}
\label{eq:dphx}
d^4 P(K)=\frac{d^4 K}{(2\pi)^4} \int d^4 x  \ \rm{exp} [
iK\cdot x -\sum_{\bf k}  {\bar n}_{\bf k} (1-e^{-i k\cdot x})
]
\end{equation}
In this formulation, an important property of the integrand in Eq.~(\ref{eq:dphx}) is that by its definition $d^4 P(K)\ne 0$ only for $K_0=\omega \ge 0$.

If one takes  the continuum limit, integrating over the three momentum $\bf K$ leads to the  probability of finding an  energy loss in the interval $d\omega$ as
\begin{equation}
\label{eq:dpomega}
dP(\omega)=\frac{d\omega}{2\pi}\int_{-\infty}^{+\infty} dt \ \rm{exp[i\omega \ t  - \beta \int_o^{\epsilon}\frac{dk}{k}(1-e^{-ik t}) ] }=\frac{d\omega}{2\pi}\int_{-\infty}^{+\infty} dt \ \rm{exp[i\omega \ t  - h(t)] }
\end{equation}
where $\beta$ is a function of the incoming and outgoing particle momenta,  and $\epsilon $ an energy scale valid for single photon emission, to be determined to the order of precision in the perturbation treatment of the process under examination. In \cite{Etim:1967} the function $\beta$ was shown to be  a relativistic invariant, and its expression  in terms of the Mandelstam variables $s,t,u$ can be found in \cite{PANCHERISRIVASTAVA1973109}.

We postpone   the implication of taking this continuum limit  to Sect.~\ref{sec:Palumbo},  and pass to evaluate Eq.~(\ref{eq:dpomega}) in closed form for $\epsilon/\omega >1$. 
Following the steps taken from Eq.~[11] through Eq.~ [17] of \cite{Etim:1967}, the analiticity properties of $h(t)$ in the lower half of the $t$-plane lead to
\begin{equation}
\label{eq:dpomegaclose}
N(\beta) dP(\omega)= \beta \frac{d\omega}{\omega} (\frac{\omega}{\epsilon})^{\beta} \ \ \ \ \  for \  \omega< \epsilon
\end{equation}
with the normalization factor given by 
\begin{equation}
\label{eq:beta}
N(\beta)=\frac{\int_o^\infty dP(\omega)}{\int_o^\epsilon dP(\omega)} = \gamma^\beta \Gamma(1+\beta)
\end{equation}
which one obtains following the procedure outlined in Appendix III of \cite{Etim:1967} and which corresponds to the results  in \cite{Lomon:1956,Lomon:1959,Erikson:1961}. This approach is based on a separation of soft from hard processes in the observed cross-section.  The application of    Touschek's method to  the scattering amplitudes  led  to the  coherent state approach to infrared effects proposed by  Greco and Rossi,   where the factor of Eq.~(\ref{eq:beta})  also appears  \cite{Greco:1967zza}. 
%and its  \tr{should we postpone this tothe ackno.?}

The question of the scale $\epsilon$ in resummation procedures acquires a particular relevance when the   matrix element of the hard scattering process  has a strong dependence on energy. In such case a straightforward   separation of the soft photon factor from  the cross-section cannot be done. Lomon considered this possibility in \cite{Lomon:1956}, when he  proposed the use of an expression such as 
\begin{equation}
\label{eq:LomonRes}
M=\int {\mathcal B}(K) Q(p-K) dK 
\end{equation}
where ${\mathcal B}(K) $ is the matrix element for the soft radiation component with total momentum $K$.  Not so relevant phenomenologically in 1955 (when Lomon submitted his paper, from the Institut of Theoretish Fysik in  Copenhagen),  the extension of the radiative correction calculation with  Eq.~(\ref{eq:LomonRes}) to processes in which a narrow resonance is produced,  was considered in \cite{Etim:1966zz,Pancheri:1969yx} and became essential  in 1974 with the  discovery \cite{Aubert:1974js,Augustin:1974xw,Bacci:1974za} of a very narrow resonance, to be called  $J/\Psi  $,  a bound state of a new type of quarks, the charm \cite{Glashow:1970gm}. 
{In \cite{Greco:1975rm} the application of the methods inspired by Lomon's expression of Eq.~(\ref{eq:LomonRes})  led to  what was, at the time, the most precise  determination of the  $J/\Psi$ width.
  In this paper, it was  found  that for very narrow resonances   the scale which controls radiative effects is not  the experimental resolution $\Delta E$ \cite{Yennie:1974ga}, but, most importantly, the resonance width  $\Gamma\le \Delta E$.}

\section{The zero momentum mode of abelian gauge fields} 
\label{sec:Palumbo}

In this section we  return to Eq.~(\ref{eq:dphx}) and  discuss  the  separation of  the zero momentum mode   from the continuum in Abelian gauge theories in the presence of  different  boundary conditions \cite{Palumbo:1983aa}.

Up to Eq.~(\ref{eq:dphx}), the method developed to obtain the energy-momentum distribution $K_\mu$ is a  classical statistical mechanics exercise.   Going further requires to input an  expression for the average number of photons of momentum $\bf k$ and the choice of the boundary  conditions imposed upon the field. To take the continuum limit, let  the quantization volume be  $V=L^3$, and introduce $\mu$, a fictitious photon mass. We must eventually take the limit $L \rightarrow \infty $ and $\mu \rightarrow 0$.  Separating the zero momentum mode of energy $\omega_0$  from all the other modes, we write
\begin{equation}
\label{eq:hdt}
h(t)=n_0(t) [1-e^{-i\omega_0 t}] + {\bar h}(t)= n_0(t) [1-e^{-i\omega_0 t}]  + \beta \int _0^E \frac{dk}{k}[1-e^{-i kt}]
\end{equation}
with the photon mass now safely taken to be zero in the integral and 
\begin{equation}
\beta=\frac{\alpha}{(2\pi)^2} \int d^2 {\bf n} \sum_{\hat e} 
| \sum_i   
\frac{
(p_i\cdot {\hat e}) \epsilon_i
}
{
({\bf p}_i \cdot {\hat n} - p_{0i})}|^2
%%+ (\frac{L}{2\pi})^3 \int d^3 {\bf k} { \bar n}_{\bf k} (1-e^{-i\omega_k t})
\end{equation}
where $p_i$ and ${\hat e}$ are  the 4-momenta and polarization of the  incoming and outgoing particles, $\epsilon_i=\pm 1$, for incoming particles or antiparticles; E the maximum single photon energy characterizing the process,  to be determined through  the inclusion of higher order corrections and the theoretical precision required for the calculation.

For the zero mode, the $\mu \rightarrow 0$ limit  is more delicate. One has
\begin{equation}
n_0(t) [1-e^{-i\omega_0 t}] \approx i W_0 t
\end{equation}
with $W_0=\frac{2\pi e^2}{L^3 \mu^2}|\sum_i \epsilon_i {\bf v_i} |^2$.
In general, one can then write
\begin{equation}
d {\mathcal P}(K_0)=\frac{1}{N(\beta)}\beta
%\gamma^\beta \Gamma(1+\beta)} \beta
\frac{dK_0}{2\pi}\int dt e^{i (K_0 - W_0)t -{\bar h}(t)}=\frac{1}{N(\beta)} \beta \frac{d K_0}{(K_0-W_0)}(\frac{K_0-W_0}{E})^{\beta} \Theta(K_0-W_0)
\end{equation}
 If one  takes first  the limit $L\rightarrow \infty$, $W_0=0$  and the zero mode gives zero contribution, as in the case of vanishing boundary conditions. The case $W_0\ne 0$ on the other hand might be present in theories with a different infrared regularization scheme. One cannot exclude  the zero mode to be relevant in the discussion of   the still unknown infrared behaviour of QCD, or in cosmology.

%In general, one can then write
%\begin{equation}
%d {\mathcal P}(K_0)=\frac{1}{N}\beta
%\gamma^\beta \Gamma(1+\beta)} \beta
%\frac{dK_0}{2\pi}\int dt e^{i (K_0 - W_0)t -{\bar h}(t)}=\frac{1}{N} \beta \frac{d K_0}{K_0}(\frac{K_0-W_0}{E})^{\beta} \Theta(K_0-W_0)
%\end{equation}
%While in the QED case $W_0=0$, one cannot exclude  the zero mode to be relevant in the discussion of   the still unknown infrared behaviour of QCD.

\section{Conclusions and Acknowlegments}
%\noindent \tr{\bf optional :We acknowledge recent  useful discussions with Fabrizio Palumbo concerning the zero momentum mode and with Mario Greco about infrared resummation methods.}  
We have shown how  Earle Lomon's    work  of the 1950's about infrared radiative corrections \cite{Lomon:1956,Lomon:1959} took the way of  Frascati, where ADONE, a 3 GeV c.m. electron-positron collider, was being built. The need to `administer'    such corrections in order to extract  meaningful physics from future experimental measurements, was keenly felt by Bruno Touschek, who had proposed and built the first electron-positron collider, AdA 
%\cite{Bernardini1964}.
\cite{Bernardini:1960osh,Bernardini:1964lqa}. 
Through  the  1960's,  Earle's work    influenced  Bruno Touschek to develop a method for infrared photon resummation, 
%later amply used in early collider experiments, such as at ACO and Novosibirsk. This is 
a legacy which Touschek  passed on  to the young theorists of the Frascati theory group, and which, along the years, can still be found  in many  extensions  to QCD  
%and application to  transverse momentum distributions  in QCD  
\cite{Parisi:1979se,Greco:2019}.

We  are   indebted  to Earle Lomon for advice and suggestions through the years we have known each other. One of us in particular, G.P., is  grateful to Earle for very many enlightening  physics conversations held together at the CTP, the MIT Center for Theoretical Physics,
spanning several years. 

G.P.  gratefully acknowledges the hospitality  at    MIT CTP 
  for  current and past  research,  and for kind   administrative support.

\bibliographystyle{phaip}
%plainnat}
%unsrtnat
%aipauth4-1}
%apsrev4-1}
\bibliography{Touschek_book_resummation_11nov2020}

\begin{thebibliography}{10}

\bibitem{Lomon:1956}
E.~L. Lomon,
\newblock Nuclear Physics {\bf 1}, 101  (1956).

\bibitem{Lomon:1959}
E.~L. Lomon,
\newblock Phys. Rev. {\bf 113}, 726 (1959).

\bibitem{Amaldi:1981}
E.~Amaldi,
\newblock {\em {The Bruno Touschek legacy (Vienna 1921 - Innsbruck 1978)}},
\newblock Number 81-19 in CERN Yellow Reports: Monographs, CERN, Geneva, 1981.

\bibitem{Bloch:1937pw}
F.~Bloch and A.~Nordsieck,
\newblock Physical Review {\bf 52}, 54 (1937).

\bibitem{Schwinger:1949}
J.~Schwinger,
\newblock Phys. Rev. {\bf 75}, 898 (1949).

\bibitem{Brown:1952}
L.~M. Brown and R.~P. Feynman,
\newblock Phys. Rev. {\bf 85}, 231 (1952).

\bibitem{Jauch:1954}
J.~M. Jauch and F.~Rohrlich,
\newblock Helvetica Physics Acta {\bf 27}, 613 (1954).

\bibitem{Jauch:1955}
J.~M. Jauch and F.~Rohrlich,
\newblock {\em {The Theory of Photons and Electrons}},
\newblock Addison-Wesley Educational Publishers Inc, 1955.

\bibitem{Yennie:1957}
D.~R. Yennie and H.~Suura,
\newblock Phys. Rev. {\bf 105}, 1378 (1957).

\bibitem{Yennie:1961}
D.~R. {Yennie}, S.~C. {Frautschi}, and H.~{Suura},
\newblock Annals of Physics {\bf 13}, 379 (1961).

\bibitem{Etim:1966zz}
E.~Etim and B.~Touschek,
\newblock {A PROPOSAL FOR THE ADMINISTRATION OF RADIATIVE CORRECTIONS,
  LNF-66-10}, 1966.

\bibitem{Etim:1967}
E.~Etim, G.~Pancheri, and B.~Touschek,
\newblock Il Nuovo Cimento B {\bf 51}, 276 (1967).

\bibitem{PANCHERISRIVASTAVA1973109}
G.~Pancheri-Srivastava,
\newblock Physics Letters B {\bf 44}, 109  (1973).

\bibitem{Erikson:1961}
K.~E. Eriksson,
\newblock Il Nuovo Cimento (1955-1965) {\bf 19}, 1010 (1961).

\bibitem{Greco:1967zza}
M.~Greco and G.~Rossi,
\newblock Nuovo Cim. {\bf 50}, 168 (1967).

\bibitem{Pancheri:1969yx}
G.~Pancheri,
\newblock Nuovo Cim. {\bf A60}, 321 (1969).

\bibitem{Aubert:1974js}
J.~J. Aubert et~al.,
\newblock Phys. Rev. Lett. {\bf 33}, 1404 (1974).

\bibitem{Augustin:1974xw}
J.~E. Augustin et~al.,
\newblock Phys. Rev. Lett. {\bf 33}, 1406 (1974),
\newblock [Adv. Exp. Phys.5,141(1976)].

\bibitem{Bacci:1974za}
C.~Bacci et~al.,
\newblock Phys. Rev. Lett. {\bf 33}, 1408 (1974),
\newblock [Erratum: Phys. Rev. Lett.33,1649(1974)].

\bibitem{Glashow:1970gm}
S.~L. Glashow, J.~Iliopoulos, and L.~Maiani,
\newblock Phys. Rev. {\bf D2}, 1285 (1970).

\bibitem{Greco:1975rm}
M.~Greco, G.~Pancheri-Srivastava, and Y.~Srivastava,
\newblock Nucl. Phys. {\bf B101}, 234 (1975).

\bibitem{Yennie:1974ga}
D.~R. Yennie,
\newblock Phys. Rev. Lett. {\bf 34}, 239 (1975).

\bibitem{Palumbo:1983aa}
F.~Palumbo and G.~Pancheri,
\newblock Phys. Lett. B {\bf 137}, 401 (1984).

\bibitem{Bernardini:1960osh}
C.~Bernardini, G.~F. Corazza, G.~Ghigo, and B.~Touschek,
\newblock Il Nuovo Cimento {\bf 18}, 1293 (1960).

\bibitem{Bernardini:1964lqa}
C.~Bernardini et~al.,
\newblock Il Nuovo Cimento {\bf 34}, 1473 (1964).

\bibitem{Parisi:1979se}
G.~Parisi and R.~Petronzio,
\newblock Nucl. Phys. {\bf B154}, 427 (1979).

\bibitem{Greco:2019}
M.~Greco,
\newblock {\em {Coherent States in Gauge Theories and Applications in Collider
  Physics}},
\newblock World Scientific, Singapore, 2019.

\end{thebibliography}
\end{document}